\documentclass[nohyper,12pt,letterpaper]{JHEP}
\usepackage[dvips]{epsfig}
\usepackage{epsf,amsfonts,amssymb}

\newcommand{\eqn}[1]{(\ref{#1})}
\newcommand{\be}{\begin{equation}}
\newcommand{\ee}{\end{equation}}
\newcommand{\ben}{\begin{displaymath}}
\newcommand{\een}{\end{displaymath}}
\newcommand{\bea}{\begin{eqnarray}}
\newcommand{\eea}{\end{eqnarray}}
\newcommand{\bean}{\begin{eqnarray*}}
\newcommand{\eean}{\end{eqnarray*}}
\newcommand{\nn}{\nonumber \\}
\newcommand{\ba}{\begin{array}}
\newcommand{\ea}{\end{array}}
\newcommand{\bi}{\begin{itemize}}
\newcommand{\ei}{\end{itemize}}

\newcommand{\epjd}[3]{{\it Eur. Phys. J. Direct.} {\bf C #1} {(#2)} #3}

\newbox\SlashedBox
\def\fs#1{\setbox\SlashedBox=\hbox{#1}
\hbox to
0pt{\hbox to 1\wd\SlashedBox{\hfil/\hfil}\hss}{#1}}
\def\hboxtosizeof#1#2{\setbox\SlashedBox=\hbox{#1}
\hbox to
1\wd\SlashedBox{#2}}

\def\ms#1{\setbox\SlashedBox=\hbox{$#1$}
\hbox to 0pt{\hbox to
1\wd\SlashedBox{\hfil/\hfil}\hss}#1}

\newcommand{\Bsm}{\,{\raisebox{1pt}{$/$} \hspace{-8pt} B}}

\def\a {\alpha}
\def\b {\beta}
\def\g {\gamma}
\def\d {\delta}

\def\G {\Gamma}
\def\e {\epsilon}

\renewcommand{\O}{\Omega}

\renewcommand{\t}{\theta}

\newcommand{\ithree}{{\it 3}}
\newcommand{\ifour}{{\it 4}}
\newcommand{\ifive}{{\it 5}}

\newcommand{\iten}{{\it 10}}

\newcommand{\caln}{\mbox{${\cal N}$}}
\newcommand{\calo}{\mbox{${\cal O}$}}

\newcommand{\calw}{\mbox{${\cal W}$}}

\newcommand{\bfe}{\mbox{\boldmath $E$}}

\newcommand{\bfna}{\mbox{\boldmath $\nabla$}}

\newcommand{\bfx}{\mbox{\boldmath $x$}}
\newcommand{\bfy}{\mbox{\boldmath $y$}}

\newcommand{\bfz}{\mbox{\boldmath $z$}}

\newcommand{\mbf}[1]{\mbox{\boldmath ${#1}$}}

\newcommand{\bbe}[1]{{\mathbb E}^{#1}}
\newcommand{\bbr}[1]{{\mathbb R}^{#1}}

\newcommand{\order}[1]{\calo\left(#1\right)}

\newcommand{\pa}{\partial}
\newcommand{\fc}{\frac}
\newcommand{\w}{\wedge}
\newcommand{\sac}{\, , \qquad}

\newcommand{\ra}{\rightarrow}

\newcommand{\adss}[2]{$AdS_{#1} \times S^{#2}$}

\newcommand{\sect}[1]{\setcounter{equation}{0}\section{#1}}
\renewcommand{\theequation}{\arabic{section}.\arabic{equation}}

\title{Penrose Limits of the Baryonic D5-brane}

\author{David Mateos and Selena Ng \\
   Department of Applied Mathematics and Theoretical Physics\\
   Centre for Mathematical Sciences \\
   Wilberforce Road, Cambridge CB3 0WA, United Kingdom \\
E-mail: \email{D.Mateos@damtp.cam.ac.uk, S.K.L.Ng@damtp.cam.ac.uk}}

\abstract{
The Penrose limits of a D5-brane wrapped on the sphere of 
\adss{5}{5} and connected to the boundary by $N$ fundamental 
strings, which is dual to the baryon vertex of the $\caln=4$ $SU(N)$ 
super Yang-Mills theory, are investigated. It is shown that, for null 
geodesics that lead to the maximally supersymmetric Hpp-wave 
background, the resulting D5-brane is a 1/2-supersymmetric
null brane. For an appropriate choice of radial geodesic, however, 
the limiting configuration is 1/4-supersymmetric and closely related 
to the Penrose limit of a flat space BIon.}


\preprint{DAMTP-2002-60 \\ \tt{hep-th/0205291}}

\begin{document}

\sect{Introduction}
\label{intro}

The Penrose limit of \adss{5}{5} with $N$ units of
Ramond-Ramond (RR) five-form flux is either the maximally 
supersymmetric type IIB Hpp-wave \cite{BFHP01} or ten-dimensional 
Minkowski space, depending on whether or not the vector tangent to the 
null geodesic along which the limit is taken possesses  
a component along the five-sphere \cite{BFHP02}. 
The perturbative spectrum of the type IIB superstring is exactly 
computable both in flat space \cite{GSW87} and in the Hpp-wave background 
\cite{Metsaev01}. The precise identification of the sector of the
dual $\caln = 4$ $SU(N)$ super Yang-Mills (SYM) theory that is 
selected by the Penrose limit allowed the authors of
\cite{BMN02} to reproduce these {\it perturbative} 
string spectra from the large-$N$ limit of the gauge theory.

According to the AdS/CFT correspondence in its strongest form, the
$\caln = 4$ SYM theory encodes the entire {\it non-perturbative} 
type IIB string theory in \adss{5}{5}.
In particular, D-brane states in \adss{5}{5}, 
and their excitations, should be `contained' within the field 
theory (possibly in the presence of external sources or defects).
Examples of such states include D3-brane giant gravitons 
\cite{MST00}, baryonic D5-branes \cite{Witten98}, and 
D5-branes embedded as \adss{4}{2} submanifolds (and their 
generalizations \cite{ST02}) relevant to the AdS/dCFT 
correspondence \cite{KR01}. 
An exact description of these states and their excitations 
in \adss{5}{5} is of course difficult because of the 
well-known problems for the quantization of strings in this 
background. Since this difficulty is absent once
the Penrose limit is taken, one might expect that in this 
limit the above D-brane states become exactly tractable. 
This expectation has been recently realized for the D3-brane 
giant gravitons \cite{BHLN02} and for the $AdS \times S$
D-branes \cite{ST02}. The purpose of this paper is
to determine the Penrose limit of the baryonic D5-brane. 

In terms of string theory in \adss{5}{5}, a static external 
`quark' in the SYM theory (that is, a static external 
electric charge transforming in the fundamental representation 
of $SU(N)$) is represented by the endpoint of a fundamental string 
that terminates at the $AdS$ boundary\footnote{
A dyon with electric and magnetic charges $p$ and $q$, respectively,
is represented by the endpoint of a $(p,q)$ string.} 
\cite{RY98,Maldacena98}. Similarly, a static baryon vertex, 
namely, a static gauge-invariant antisymmetric combination of
$N$ external electric charges, is represented by a D5-brane
wrapped on $S^5$ and with constant position in a constant-time slice
of $AdS_5$, and connected to the $AdS_5$ boundary by $N$ 
fundamental strings \cite{Witten98}. The reason for this is that
the flux of the background RR five-form 
through the $S^5$ sources the Born-Infeld (BI) $U(1)$ gauge field on the
D5-brane, inducing $N$ units of electric charge.
Since the D5 worldspace is compact, the
total net charge must vanish. The extra contribution is provided by
the endpoints of $N$ fundamental strings (with suitable orientations)
connecting the D5-brane to the $AdS$ boundary. 
The antisymmetry of the baryon vertex follows from the fermionic 
nature of the ground state of each string.

Each string connecting the D5-brane to the $AdS$ boundary defines a
`direction' in \adss{5}{5}. Since the supersymmetries 
preserved by a string depend on its orientation, one
might expect that a generic baryon vertex will break
supersymmetry completely. On the other hand, it might be 
suspected (correctly) that the most symmetric configuration, 
in which all strings start at the same point on the D5-brane 
and end at the same point on the $AdS$ boundary, preserves 
some fraction of supersymmetry. This is a very
natural configuration to consider also from the 
gauge theory viewpoint, since it corresponds to a 
`localized' baryon vertex, that is, one for which all the $N$ 
external quarks lie at the same point.
Despite its higher degree of symmetry, the description of this 
configuration in terms of a perfectly spherical D5-brane is 
oversimplified because it ignores (at least) two essential 
effects caused by the coincidence of all the endpoints of the 
strings on the D5:\footnote{As was realized \cite{Imamura98b} after
some early computations of the baryon mass \cite{BISY98,Imamura98a}.}
the deformation of the shape of the D5-brane by the tension of the 
strings, and the highly inhomogeneous BI electric field on the D5-brane
sourced by their endpoints. 

Fortunately, these two effects {\it are} captured by the 
Dirac-Born-Infeld (DBI) action of the D5-brane 
coupled to the \adss{5}{5} background 
\cite{Imamura98b, CGS98, CGMV99, GRST99}. 
In this description, moreover, the strings do not need to
be treated separately, but instead arise as a `spike' deformation
of the D5-brane worldvolume; in this sense, this 
configuration provides an $AdS$
analog of the flat space BIon \cite{CM97,Gibbons97}.
Since all the strings end at the same point on the D5-brane, 
an $SO(5)$ subgroup of the $SO(6)$ isometry group of the 
five-sphere is preserved. It follows that the deformation of the
D5-brane can be specified by giving the radial position of the brane
in $AdS_5$ as a function $r(\t)$ of the colatitude angle $\t$ on the
$S^5$, measured with respect to the endpoint of the strings. A
condition on the function $r(\t)$ for preservation of some
supersymmetry was first found in \cite{Imamura98b}. An explicit
solution of this condition, which is given in \eqn{r}, was presented in
\cite{CGS98}. It was subsequently shown that this
solution saturates an energy bound and hence minimizes the energy for
a fixed value of a topological charge \cite{CGMV99}. Finally, it was
verified that the fraction of preserved supersymmetry is 1/4 by making
use of the $\kappa$-symmetry transformations of the D5-brane 
\cite{GRST99}.

Since the $SO(5)$-symmetric baryonic D5-brane completely wraps an
$S^4$ inside the $S^5$, one might expect that the Penrose limit
along an appropriate null geodesic tangent to the sphere would lead to a
D5-brane extending along both light-cone directions $x^\pm$ (as well
as four other spacelike directions) in the resulting Hpp-wave  
spacetime. We will show, however, that this is not the case, the
reason being that there is no null geodesic contained within 
the D5-brane worldvolume. Consequently, in the Penrose limit the 
brane is always `pushed' to lie at constant $x^+$ 
in the metric \eqn{pp-metric}; this implies that it becomes 
a null brane, by which we mean a brane whose
worldvolume is a null hypersurface\footnote{This implies $\det g=0$,
where $g$ is the induced metric on the brane; some authors use the
term `null brane' to refer instead to branes for which 
$\det(g+F)=0$, where $F$ is the BI field strength.}.

We will also consider the Penrose limit along null geodesics in 
the $AdS$ radial direction. For generic geodesics of this type, the
resulting configuration is a null flat D5-brane in Minkowski space. 
For the radial geodesic along the strings attached to the D5-brane,
however, the Penrose limit is closely related to that of the flat 
space BIon.

\sect{Baryonic D5-branes}

We begin by briefly reviewing the relevant results from 
\cite{Imamura98b, CGS98, CGMV99, GRST99}; we refer the reader to these
references for details.

We write the metric on \adss{5}{5} as
\be
ds^2 = r^2 \left( -dt^2 + \mbf{dx}^2 \right) +
\fc{dr^2}{r^2} +  d\t^2 + \sin^2 \t \, d\Omega_\ifour^2  \,,
\label{ads-metric}
\ee
where $d\Omega_\ifour^2$ is the line element on a unit four-sphere and 
$\bfx$  are Cartesian coordinates on $\bbe{3}$. The baryonic 
D5-brane is wrapped on the $S^4$ and lies at rest at a
point in $\bbe{3}$; by translational invariance we take it to be 
\be
\bfx = 0 \,. 
\label{x}
\ee
The radial position $r$ of the D5-brane in $AdS_5$ is then specified 
\cite{CGS98} as a function of the colatitude angle $\t$ by 
\be
r(\t) = A \, f(\theta) \sac f(\theta) = 
\fc{\left( \pi - \t + \sin \t \, \cos \t \right)^{1/3}}{\sin \t} \,,
\label{r}
\ee
where $A$ is an arbitrary constant that sets the overall scale of the
solution; the freedom of changing $A$ arises from the conformal
invariance of the $AdS$ background. The profile $r(\t)$ reflects the
finite deformation of the shape of the D5-brane caused by the tension
of the strings; in particular, $r(\t)$ diverges as 
$r(\t) \sim \t^{-1}$ for $\t \ra 0$. This `spike' singularity 
represents the $N$ fundamental strings connecting the D5-brane to the
$AdS$ boundary at $r\ra \infty$.

Note that we are only considering here the
case where the integration constant $\nu$ of \cite{CGS98} is set to
zero, since it is this case which genuinely corresponds to the baryon
vertex in the dual $\caln=4$ $SU(N)$ gauge theory; the physical
interpretation of the solutions with $\nu \neq 0$ is rather unclear
\cite{CGS98}.

In addition to deforming the brane, the endpoints of the strings
source the electric components of the worldvolume BI two-form, which
thus takes the form \cite{Imamura98b,CGS98}
\be
F = A \, E(\t) \, dt \w d\t \, ,
\label{F}
\ee
where
\be
E(\t) = g(\t) \, \sqrt{\fc{f^2(\t) + f'\,^2(\t)}
{g^2(\t) + \sin^8 (\t)}} \sac 
g(\t) = \fc{3}{2} \, \left( \pi - \t + \sin \t \cos \t \right)
+ \sin^3 \t \cos \t \,.
\label{e}
\ee
Note that $F$ is $SO(5)$-invariant.

\sect{Null D5-branes in Hpp-waves}

The Penrose limit of any spacetime is a pp-wave spacetime
\cite{Penrose76, Guven00} whose metric in Brinkman coordinates takes
the form
\be
ds^2 = -4 \, dx^+ dx^- + A_{ij}(x^+) \, z^i z^j \,(dx^+)^2 
+ ds^2\left( \bbe{8} \right) \,,
\label{pp-metric}
\ee
where $z^i$ are Cartesian coordinates on $\bbe{8}$. The maximally
supersymmetric Hpp-wave solution of type IIB supergravity corresponds
to the particular case $A_{ij}=-\d_{ij}$; there is also a  
RR five-form given by
\be
F_\ifive = 4\, dx^+ \w \left( dz^1 \w \ldots \w dz^4 +
dz^5 \w \ldots \w dz^8 \right) \,. 
\ee
The Hpp-wave solution can be obtained as the Penrose limit of the 
$AdS_5 \times S^5$ solution along a null geodesic whose tangent vector 
has a non-zero projection onto $S^5$ \cite{BFHP02}; following \cite{BFP02}, 
we shall refer to such geodesics as `generic null geodesics'. 

In this section we will show that the Penrose limit of the baryonic
D5-brane along {\it any} generic null geodesic is necessarily a null 
D5-brane, that is, a D5-brane whose worldvolume is a null 
six-dimensional hypersurface in the above metric. This follows from
the general fact mentioned in the Introduction: a brane 
is always `pushed' to lie at constant $x^+$ in the pp-wave spacetime  
when the Penrose limit is taken along a null geodesic that 
intersects, but that is not contained within, its worldvolume. 
Although this result might be intuitive, we provide here a somewhat
formal justification.

In the neighbourhood of a segment of a null geodesic $\g$ containing
no conjugate points there exist local coordinates $\{U, V, Y^i\}$ 
such that the metric takes the form \cite{Penrose76, Guven00, BFP02}
\be
ds^2 = 2 dV \left( dU + \a \, dV + \b_i \, dY^i \right) +
C_{ij} \, dY^i dY^j \,,
\ee
where $\a$, $\b_i$ and $C_{ij}$ are functions of
all the coordinates. The geodesic $\g$ lies at $V=Y^i=0$ and is
affinely parametrized by $U$. The resulting metric 
$\bar{ds}^2$ in the Penrose limit along $\g$ may be 
obtained by rescaling the coordinates as 
\be
U=u \sac V=\Omega^2 \, v \sac Y^i = \Omega \, y^i  
\label{rescaling}
\ee
for positive real constant $\Omega$, and computing the limit
\be
\bar{ds}^2 \equiv \lim_{\Omega \ra 0} \Omega^{-2} \, ds^2 
\label{limit}
\ee
while keeping $u$, $v$ and $y^i$ fixed. It can then be brought
to the form \eqn{pp-metric} by a further change of variables that, in
particular, identifies $u=x^+$. 

Consider now a brane embedded with $(p+1)$-dimensional 
worldvolume $\calw$ in the spacetime in question. 
Unless $\g$ intersects $\calw$,\footnote{
Or becomes arbitrarily close to it asymptotically; see next section.}
the brane `disappears' upon taking the Penrose limit, because
only a region infinitesimally close to $\g$ (which is magnified by an
`infinite' amount) survives the limit \cite{BFP02}. Let us therefore
focus on the other two possibilities, that is, either $\g$
intersects $\calw$ at an isolated point $q=(U_0,0,0)$ or $\g$ is
actually contained within $\calw$.\footnote{
This is a local analysis; globally, of course, $\g$ might intersect 
$\calw$ at more than one point and/or have a number of
segments contained within it.} In both cases, in a neighbourhood of
$q$ the brane embedding may be specified in implicit form as 
\be
f^k(U,V,Y^i) = 0
\label{fk}
\ee
for functions $f^k$, $k=0, \ldots , p$. After the rescaling 
\eqn{rescaling} we can expand these functions as
\be
f^k(U,V,Y^i) = f^k(u, 0, 0) + 
\Omega^2 \, f^k_V(u,0,0)\, v  + \Omega \, f^k_i (u,0,0) \, y^i 
+ \cdots  \,,
\ee
where $f^k_V = \pa_V f^k$ and $f^k_i = \pa_{Y^i} f^k$.
Since $\g$ lies at $V=Y^i=0$, if $\g$ is contained within $\calw$ then 
the first term in this expansion vanishes identically 
{\it for all $u$}, and hence in the limit $\Omega \ra 0$ equations 
\eqn{fk} determine a subset of the $v$ and $y^i$ coordinates as
functions of the remaining ones and/or of $u$. 
Instead, if $\g$ is not contained in
$\calw$ then in the limit $\Omega \ra 0$ the above equations imply 
$f^k(u, 0, 0)=0$, and hence that the brane lies at $u=U_0$ or,
equivalently, at constant $x^+$ in Brinkman coordinates. 

We now return to the baryonic D5-brane. Since we wish to prove that
there is no generic null geodesic contained within the D5-brane
worldvolume, we need only consider geodesics that lie at 
$\bfx=0$. If we view $S^5$ as embedded in $\bbr{6}$ in the
usual way, then the projection on $S^5$ of any generic null
geodesic is a circle of unit radius on some two-plane. In the
three-dimensional subspace of $\bbr{6}$ generated by this plane
and the axis with respect to which the azimuthal angle $\t$ is 
measured\footnote{
If this axis is contained within the two-plane it suffices to set
$\Theta=\t$ in the discussion below.} 
we may define new spherical coordinates $\Theta, \Phi$ 
such that the geodesic lies at $\Phi=0$ and is parametrized by the new
azimuthal angle $\Theta$ (see Figure \ref{sphere}). 
\FIGURE[t]{
\epsfig{file=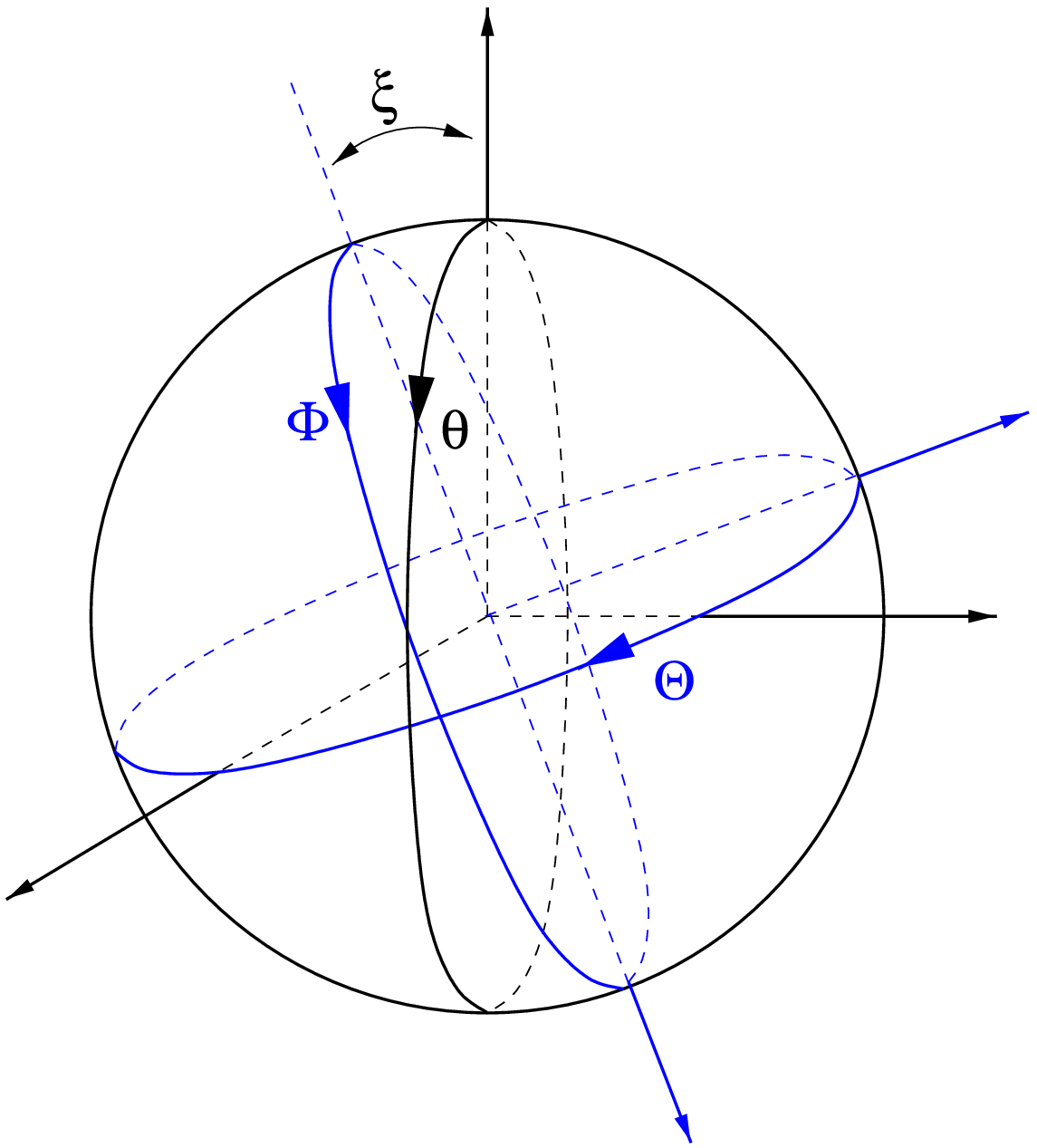, height=10cm}
\caption{New coordinates on the five-sphere.}
\label{sphere}
}
In terms of these new coordinates we have
\be
\t=h(\Theta,\Phi) \equiv \arccos \left( 
\sin \xi \cos \Theta - \cos \xi \sin \Theta \sin \Phi \right) \,,
\label{h}
\ee
where $\xi$ is the angle between the normal to the plane and the
$\t$-axis. The radial position in $AdS_5$ of the geodesic is 
given in terms of the new azimuthal angle by $r=\ell^{-1} \sin \Theta$, 
where $\ell$ is the angular momentum of the geodesic \cite{BFP02}. 
It follows that the condition for a point to belong both to the
geodesic and to the D5-brane worldvolume is 
\be
\ell^{-1} \sin \Theta = 
A \, f\left( h\left( \Theta, 0 \right) \right) \,.
\label{Theta}
\ee
Since this equation has at most two solutions (which occurs for appropriate 
values of $A$, $\ell$ and $\xi$), we conclude that there is no generic
null geodesic contained within the D5-brane worldvolume, and hence 
that the Penrose limit along any such geodesic leads to a null
D5-brane in the IIB Hpp-wave background. 

In order to determine the resulting D5-brane more precisely, we first 
change coordinates from $\{t,r,\Theta\}$ to $\{U, V, \Psi\}$ 
adapted to the geodesic and defined by \cite{BFP02}
\be
t = - \ell \cot \ell U - V + \ell \Psi \sac
r = \ell^{-1} \sin \ell U \sac 
\Theta = \ell U + \Psi \,.
\ee
The geodesic lies at $V=\Psi=\bfx=0$, so we further rescale the
coordinates as 
\be
U = u \sac V = \O^2 v \sac \bfx = \O \tilde{\bfx} \sac
\Psi = \O \psi \sac \Phi = \O y
\ee
and take the limit \eqn{limit} with the result 
\be
\bar{ds}^2 = 2 du dv + \cos^2 \ell u \, d\psi^2 + 
\ell^{-2} \sin^2 \ell u \, d\tilde{\bfx}^2 + 
\sin^2 \ell u \, d\bfy^2 \,,
\ee
where we have set $y = |\bfy|$ for Cartesian coordinates 
$\bfy$ on $\bbe{4}$. This is just the Hpp-wave metric in Rosen
coordinates. The final transformation 
\bea
u &=& x^+ \sac 
\psi = z^1 \sec \ell x^+  \sac 
\tilde{\bfx} = \ell^2 \tilde{\bfz} \csc \ell x^+ \sac 
\bfy = \bfz \csc \ell x^+ \,, \nn
v &=& 2x^- 
- \fc{\ell}{2} \tan \ell x^+  \left( z^1 \right)^2 
+ \fc{\ell}{2} \cot \ell x^+ \left( \tilde{\bfz}^2 + \bfz^2 \right)
\eea
brings the metric to the form \eqn{pp-metric} with $A_{ij}= -\d_{ij}$,
$\tilde{\bfz}=(z^2,z^3,z^4)$ and $\bfz=(z^5, z^6, z^7, z^8)$. 

The original embedding equation \eqn{x} becomes simply
$\tilde{\bfz}=0$ in the final coordinates, 
whereas \eqn{r} yields \eqn{Theta} with $\Theta$ 
replaced by $\ell x^+$, which implies $x^+=\mbox{constant}$;\footnote{
If equation \eqn{Theta} has two solutions then the Penrose
limit results in two parallel branes at two different values of $x^+$.}
this means that the brane extends along
the directions $x^-$, $z^1$ and $\bfz$. 
The limiting BI field $\bar{F}$ is similarly determined
by first performing the above changes of coordinates and
rescalings and then computing the limit 
\be
\bar{F} = \lim_{\O \ra 0} \, \O^{-2} F \,.
\label{limitF}
\ee
The result is 
\be
\bar{F} = B \, dz^1 \w dz \,,
\ee
where $z=|\bfz|$ and $B$ is a constant that can be determined
explicitly but whose precise value will not be needed in the
following. Note that the original electric field gives rise to a
magnetic field after the Penrose limit; this can presumably 
be understood as the magnetic field seen by a 
Lorentz-transformed observer whose worldline
approaches the null geodesic. 

As anticipated, we have obtained a null D5-brane in the sense that
$\det g=0$, where $g$ is the induced metric on the brane. Since we
also have
\be
\det(g + F)=0 \,,
\label{gF}
\ee
the appropriate action to describe
the dynamics of this configuration is not the ordinary DBI action 
but rather\footnote{
The coupling to the background RR five-form vanishes because of the
D5-brane orientation.} \cite{LU97,BT98}
\be
S = \int \fc{1}{2 e} \, \det (g+F) \,,
\label{action}
\ee
where $e$ is an independent worldvolume density whose equation of
motion enforces the condition \eqn{gF}. We have verified that the
remaining field equations derived from \eqn{action} are also
satisfied (with constant $e$); this should be expected given that the 
original configuration solved the D5-brane equations in \adss{5}{5}. 
Moreover, since the original solution is supersymmetric, so should 
be the resulting configuration after the Penrose limit, as we now 
proceed to show.

The supersymmetries preserved by a D5-brane are those generated 
by Killing spinors $\e$ of the background that satisfy \cite{BKOP97}
\be
\Xi \, \e = \sqrt{-\det(g+F)} \, \e \,,
\label{susy}
\ee
where $\Xi$ is the matrix occurring in the kappa-symmetry
transformations of the fermions in the supersymmetric extension of 
the action \eqn{action} \cite{BT98}; by construction, 
\mbox{$\Xi^2 = -\det(g+F)$}, so in the present case $\Xi$ is
nilpotent. 

The background metric \eqn{pp-metric} can be written as
\be
ds_\iten^2 = 2 \, e^+ e^- + e^i e^i 
\ee
for orthonormal one-forms  
\be
e^- = 2 \, dx^- + \fc{z^i z^i}{2}\, dx^+ \sac 
e^+ = - dx^+ \sac e^i = dz^i \,.
\label{forms}
\ee
In this basis the Killing spinors take the form \cite{BFHP01,BFHP02,ST02}
\be
\e = \left( 1 - \fc{i}{2} \, \G_- M \right) \, 
\exp \left(-\fc{i}{2} \, x^+ \, N \right) \, \e_0\,,
\ee
where 
\be
M = \sum_{a=1}^4 z^a \G_a \G_{1234} + z^{a+4} \G_{a+4} \G_{5678} \sac
N = \G_{1234} + \G_{5678} \,.
\ee
Here $\{\G_+, \G_-, \G_1, \ldots, \G_8\}$ are real tangent-space 
constant Dirac matrices associated to the basis \eqn{forms}, and $\e_0$ is
a complex 16-component constant spinor of negative chirality, that is,
$\G_{+-12345678} \e_0 = -\e_0$. 

For the configuration of interest here the kappa-symmetry matrix takes
the form
\be
\Xi = 2 \G_{15678} \left( \G_1 \Bsm  - K \right) I \G_- \,,
\ee
where the operations $K$ and $I$ act on spinors as
\be
K \e = \e^* \sac I \e = -i \e 
\ee
and where we have defined 
\be
\Bsm = \sum_{a=5}^8 B_a \G^a \sac B_a = \fc{z^a}{z} B \,.
\ee
Equation \eqn{susy} thus becomes
\be
2 \G_{15678} \exp \left( -\fc{i}{2} \, x^+ \, N \right)
\left( \G_1 \Bsm - K \right) I  \G_- \e_0 = 0 \,,
\ee
where we have made use of the fact that $\G_-^2=0$. Since the matrix 
acting on $\G_- \e_0$ is invertible, we conclude that the only
solutions to this equation are provided by spinors that satisfy
\be
\G_- \e_0 = 0 \,.
\ee
This means that 1/2 of the 32 supersymmetries of the Hpp-wave
background are unbroken by the D5-brane, which is twice the amount of
supercharges preserved by the original baryonic D5-brane. This
enhancement of supersymmetry in the Penrose limit has been observed 
previously in \cite{IKM02, GO02, PS02}. Note that, unlike in
\cite{ST02}, all the D5-brane configurations in the Hpp-wave
background with the same orientation as the one we have obtained 
via the Penrose limit preserve the same supersymmetries,
irrespectively of whether or not the values of $x^+$ and $B$ 
are those particular ones dictated by the limit.

\sect{Radial Geodesics and D5-brane BIons} 

In this section we shall examine the Penrose limit of the baryonic
D5-brane along radial null geodesics, by which we mean 
null geodesics that lie at a point $p$ on $S^5$ and at $\bfx=0$ in
$AdS_5$. If $p$ is not the North pole, that is, the point $\t=0$, then 
the radial geodesic intersects the D5-brane at one point and the
Penrose limit along it leads to a null D5-brane in Minkowski
space; since this is similar to the case discussed in detail in the
previous section, we shall not elaborate on this further. Instead, 
in this section we shall focus on the Penrose limit along the geodesic at 
$\t=0$.  Since this does not intersect the D5-brane, 
it might seem that the D5-brane disappears upon taking the 
limit. However, because this geodesic becomes arbitrarily 
close to a region of the brane worldvolume asymptotically 
(that is, as $r \ra \infty$), this region survives the limit and 
leads to a D5-brane embedded non-trivially in Minkowski space.

The change of coordinates 
\be
U = r \sac V = t + r^{-1}
\label{change}
\ee
brings the metric \eqn{ads-metric} to the form
\be
ds^2 = 2 dU dV - U^2 dV^2 + U^2 \mbf{dx}\,^2 + d\t^2 +
\sin^2\t\, d\O_\ifour^2 \,.
\ee
The radial geodesic of interest lies at $V=\t=\bfx=0$, and is
parametrized by $U$. To take the Penrose limit we perform the
usual rescaling 
\be
U=u \sac V=\Omega^2 v \sac \bfx =\O \tilde{\bfx}
\sac \t = \O y
\ee
and take the limit \eqn{limit}. The resulting metric is
\be
\bar{ds}^2 = 2 du dv + u^2 \mbf{d\tilde{x}}^2  + \mbf{dy}^2 \,, 
\ee
where we have set $y=|\bfy|$ for Cartesian coordinates $\bfy$ 
on $\bbe{5}$. The further coordinate transformation
\be
u = \frac{X+y^0}{2} \sac
v = (X-y^0) + \frac{\mbf{w}^2}{X + y^0} \sac
\mbf{\tilde x}  = \frac{2 \mbf{w}}{X+y^0} 
\label{cartesian}
\ee
brings the metric to the form
\be
\bar{ds}^2 = - \left(dy^0 \right)^2 + \mbf{dy}^2 + dX^2 
+ \mbf{dw}^2 \,,
\ee
which, as expected \cite{BFP02}, is just the Minkowski metric. 

The embedding equation \eqn{x} for the D5-brane becomes simply 
$\mbf{w}=0$ in the new coordinates, while equation \eqn{r} yields
\be
X + y^0  = 2 A \, f(\O y) = \fc{2 \pi^{1/3} A}{\O y} + \order{\O} \,.
\ee
We see that, because $r(\t)$ diverges as $\t\ra 0$, in order for 
the limit $\O \ra 0$ with finite $y^0$ and $X$ to exist 
we {\it must} rescale 
\be
A = \fc{a}{2 \pi^{1/3}} \, \O
\ee
for constant, $\O$-independent $a$ (the numerical factor is for 
later convenience). Note that we are allowed to do so because 
\eqn{r} yields a (supersymmetric) solution of the D5-brane equations 
of motion {\it for any $A$}. The rescaling above means that, 
as we focus on a region closer and closer to the geodesic, 
we also focus on a solution with smaller and smaller $A$. 
This procedure gives a finite limit for the embedding equation, namely
\be
X = \frac{a}{y} - y^0 \,.
\label{X}
\ee
Thus in the Penrose limit we may interpret $\{y^0, \bfy\}$ as
worldvolume coordinates of the D5-brane and $X$ as the only excited 
transverse scalar with profile given by \eqn{X}. 

The limiting BI field strength on the D5-brane is obtained in an
analogous manner by performing the coordinate tranformations and the
rescalings above, and then computing the limit \eqn{limitF}, for which
it is useful to note that
\be
E(\O y) = \fc{\pi^{1/3}}{\O^2 \, y^2} + \order{1} \,.
\ee
The result is $\bar{F} = E_a(\bfy) \, dy^0 \w dy^a$, where
\be
\bfe = \bfna X \,. 
\label{BPS1}
\ee
Note that precisely the same rescaling of $A$ required to obtain a
well-defined limiting embedding also yields a finite BI field. The
ultimate reason behind this is presumably supersymmetry, since this
determines the original electric field in terms of the scalar 
transverse to
the D5-brane (the radial position $r$ in $AdS_5$), as specified by
\eqn{e}. This non-linear equation simplifies in the
Penrose limit to \eqn{BPS1}, which is precisely the condition 
for an ordinary Minkowski space BIon to preserve some fraction 
of supersymmetry \cite{CM97, Gibbons97, GGT97}. In fact, the Penrose 
limit of the ordinary BIon along the direction of the string 
yields a transverse scalar 
\be
X = \frac{a}{y^3} - y^0 
\label{Xbion}
\ee
and an electric field given by \eqn{BPS1} (see the Appendix for
details). Note that the only difference is that in this case $X$ is a 
harmonic function on the D5-brane {\it worldspace}, as required 
by its equation of motion before the Penrose limit. 
It might seem surprising, therefore, that both configurations
\eqn{X} and \eqn{Xbion} are supersymmetric and solve the same D5-brane 
equations of motion, as one would expect from their origin as
Penrose limits of supersymmetric solutions. We will return to this issue
below. 

Since in both cases 
\be
\dot{X} = -1 \,,
\label{BPS2}
\ee
where $\dot{X} \equiv \pa_0 X$, each point on the brane 
is moving in the $X$-direction at the speed of light. 
One might think that this is only possible for a
brane whose induced metric $g$ is null in the sense that 
$\det g=0$. However, because a
D-brane is boost-invariant, only the component of the velocity
orthogonal to its surface is physical, and this ensures that 
the physical velocity of all points on the D5-brane is 
subluminal\footnote{An analogous situation in the context of
fundamental strings has been recently discussed in \cite{MNT02}.}. 
In particular, we find
\be
\det g = -1 + \dot{X}^2 - \left(\bfna X\right)^2 \,,
\ee
and therefore the induced metric on the D5-brane is everywhere Lorentzian
(although it does approach a null metric as $y \ra \infty$). 
However, a calculation reveals that 
\be
\det (g+F) = - \left(1 - \bfe^2 \right) 
\left(1+ \left(\bfna X \right)^2 \right)  
- \left( \bfe \cdot \bfna X \right)^2 +  \dot{X}^2 
\ee
does vanish, and consequently that the appropriate action to describe
the D5-brane dynamics is again \eqn{action}. For configurations
satisfying \eqn{BPS1} and \eqn{BPS2} the equations of motion obtained
by varying this action with respect to $X$ and to the BI gauge field
reduce to the unique condition
\be
\bfna \cdot \left( e^{-1} \, \bfna X \right) = 0 \,.
\label{unique}
\ee
We thus see that the two configurations \eqn{X} and 
\eqn{Xbion} are particular cases of a more general class 
of solutions characterized by the fact that $X$ depends solely on 
the radial coordinate $y$ on the D5-brane worldspace;
these solutions satisfy equation \eqn{unique} with $e = y^4 \, dX/dy$. 
This arbitrariness in the profile of the D5-brane is reminiscent of
that recently discussed in the context of D2-brane supertubes
\cite{MNT01}.   

We finally turn to showing that the conditions \eqn{BPS1} and \eqn{BPS2} 
guarantee not only that the D5-brane field equations are 
satisfied but also that 1/4 supersymmetry is preserved. Again we must
seek solutions of equation \eqn{susy}, where now $\e$ is a 
{\it constant} 16-component complex Killing spinor of the Minkowski
background and  
\be
\Xi = \G_{012345} \, K I \, \left( 1 -\dot{X} \, \G_{0X} \right)
+ \G_{12345} \G^a \, I \, \left( E_a - \pa_a X \, \G_{0X} K \right)
+ \G_{12345X} \, I \, E_a \, \pa_b X \, \G^{ab} \,,
\ee
where $a, b=1, \ldots ,5$ and, as in the previous section, the Dirac 
matrices appearing here are real flat-space constant matrices.
It follows that if conditions \eqn{BPS1} and \eqn{BPS2} hold then
equation \eqn{susy} is verified by spinors subject to the constraints
\be
\G_{0X} K \, \e =  \e  \sac \G_{0X} \, \e = -\e \,.
\ee
The first condition is associated to fundamental strings along the
$X$-direction, whereas the second one corresponds to a wave along the same
direction. Since the two constraints are mutually compatible, 1/4
supersymmetry is preserved. In the two particular cases discussed
above, the strings in the $X$-direction arise of course from the
strings originally attached to the D5-brane before the Penrose limit. 
Instead, the wave arises because in taking the limit the brane is 
effectively boosted to the speed of light in the $X$-direction, 
as discussed above. Note that there is no trace of a constraint 
associated to the presence of a D5-brane, again in analogy 
to the case of D2-brane supertubes \cite{MT01}. 

\sect{Discussion}

We have examined in detail the possible Penrose limits of the D5-brane
dual to the baryon vertex of the $\caln =4$ SYM theory. While the
resulting background depends solely on whether or not the null geodesic along
which the limit is taken possesses a non-zero projection
on $S^5$, the limit of the baryon vertex depends on the relative
orientation between the geodesic and the D5-brane. 

For generic null geodesics, that is, for 
geodesics that lead to the maximally supersymmetric Hpp-wave 
solution, the resulting 1/2-supersymmetric D5-brane is always a 
null brane, in the sense that its worldvolume is a null-hypersurface. 
This situation should be contrasted with the Penrose 
limit of the D3-brane giant gravitons \cite{BHLN02} and of the 
$AdS \times S$ D-branes \cite{ST02}. These branes are static 
with respect to the $AdS$ {\it global} time\footnote{
Although the giant gravitons with angular momentum $J<N$ are
time-dependent configurations, the case $J=N$ considered 
in \cite{BHLN02} corresponds to a static D3-brane whose angular
momentum arises entirely from its Wess-Zumino coupling to 
the background RR five-form.} $\tau$ and wrap some great circle (or a
higher-dimensional submanifold) within the $S^5$. For such types of
branes there is always a null geodesic {\it contained} within their
worldvolumes. To see this, write the metric on \adss{5}{5} as
\be
ds^2 = - \cosh^2 \rho \, d\tau^2 + d\rho^2 + 
\sinh^2 \rho \, d\O_\ithree^2 + \cos^2 \phi \, d\psi^2 +
d\phi^2 + \sin^2 \phi \, d\O_\ithree^2 \,,
\ee
where $\psi$ is the coordinate along the great circle. Then the curve 
$\tau=\psi$, $\rho=\phi=0$ is a null geodesic (of the type 
considered in \cite{BMN02}) that clearly lies within the 
worldvolume of the brane in question. The Penrose limit along this 
geodesic leads to a brane that extends along both $x^+$ and $x^-$ 
(and possibly some other spatial directions) in the Hpp-wave metric 
\eqn{pp-metric}, and therefore the induced metric possesses 
Lorentzian signature.
In contrast, the baryonic D5-brane is static with respect to 
the {\it Poincar\'e} time $t$, and we have seen that in this case 
there is no generic null geodesic contained within its worldvolume;
for geodesics that intersect it at a point, the D5-brane is 
`pushed' to lie at constant $x^+$ in the resulting pp-wave
background. This result makes the physical interpretation 
of the baryonic D5-brane in the Penrose limit unclear, especially in 
relation to the $\caln=4$ SYM theory. We leave the investigation of
this point for the future. 

We have also analyzed the Penrose limit along radial null geodesics
that lie at a point $p$ on $S^5$. If $p$ is not the North pole, that
is, the point at which the fundamental strings end on the D5-brane, 
then the resulting configuration is a null flat D5-brane in Minkowski 
space. If $p$ is the North pole then the radial geodesic does not
intersect the D5-brane, so it might seem that the latter disappears
upon taking the Penrose limit. However, we showed that if the
overall size of the baryon vertex is appropriately rescaled then the limit 
results in a 1/4-supersymmetric D5-brane closely related to the 
Penrose limit of the flat space BIon.
\newline
\newline
{\bf Note added:} While this paper was being type-written we received 
\cite{Seki02}, where some issues related to those reported here 
are considered.

\section*{Acknowledgments}

We are grateful to Prem Kumar for collaboration at earlier stages of
this work, and to Michael B. Green for discussions.
D.M. is supported by a PPARC fellowship.  S.N. is
supported by the British Federation of Women Graduates and the
Australian Federation of University Women (Queensland).

\appendix
\sect{Penrose Limit of the Minkowski Space BIon}
\renewcommand{\theequation}{A.\arabic{equation}}

The BIon \cite{CM97, Gibbons97, GGT97} is a solution of the equations 
of motion of a D-brane embedded in Minkowski space that represents a 
semi-infinite string ending orthogonally on the D-brane; it preserves
1/2 of the D-brane worldvolume supersymmetry (that is, 1/4 of the
32 Minkowski supersymmetries).

We write the Minkowski metric as 
\be
ds^2 = -dT^2 + d\mbf{Y}^2 + dX^2 + d\mbf{Z}^2 \,,
\label{mink-metric}
\ee
where $\mbf{Y}$ and $\mbf{Z}$ are Cartesian coordinates on $\bbe{5}$
and $\bbe{3}$, respectively. Suppose the D5-brane lies at $\mbf{Z}=0$ and
that $X$ is the direction of the string. Any configuration such that
$\bfe =\bfna X$ and $\bfna^2 X=0$, where $\bfe$ is the BI electric
field and $\bfna$ is the gradient operator in $\bbe{5}$, 
solves the D5-brane field equations and preserves 1/2 supersymmetry. 
For spherically symmetric configurations we thus have 
\be
X = \fc{A}{Y^3} \,,
\label{Xappendix}
\ee
where $A$ is a constant and $Y=|\mbf{Y}|$. This represents a string
ending on the brane at $Y=0$. 

In order to take the Penrose limit along the null geodesic $T=X$ that
runs along the string we define new coordinates 
\be
U = T+X \sac V=T-X \,,
\ee
perform the rescaling 
\be
U=u \sac V=\O^2 v \sac \mbf{Y} = \O \bfy \sac  \mbf{Z} = \O \bfz \,,
\ee
and take the limit \eqn{limit}. The resulting metric
\be
\bar{ds}^2 = -du\, dv + d\bfy^2 + d\bfz^2 
\ee
is of course the Minkowski metric again. Equation \eqn{Xappendix} in the
rescaled coordinates becomes
\be
u = \frac{2A}{\O^3 y^3} + \O^2 v \,,
\ee
so we see that to achieve a finite limit we must rescale $2A = \O^3 a$
for constant $\O$-independent $a$, in which case we obtain 
\be
x = \fc{a}{y^3} - t \,,
\ee
(where we have set $u=t+x$ and $v=t-x$) thus reproducing \eqn{Xbion}. 
The resulting electric field is similarly computed by performing the
coordinate changes and rescalings above and taking the limit
\eqn{limitF}, with the expected result $\bar{\bfe} = \bfna x$. 

The configuration we have obtained can also be derived, as might be
expected, by boosting the
initial BIon along the negative $X$-direction to the speed 
of light while rescaling $A$ appropriately in order to achieve a
finite limit.




\begin{thebibliography}{80}

\bibitem{BFHP01}
M.\ Blau, J.\ Figueroa-O'Farrill, C.\ Hull and G.\ Papadopoulos,
{\it A New Maximally Supersymmetric Background of Type IIB 
Superstring Theory}, \jhep{01}{2002}{047}, \hepth{0110242}.

\bibitem{BFHP02}
M.\ Blau, J.\ Figueroa-O'Farrill, C.\ Hull and G.\ Papadopoulos,
{\it Penrose Limits and Maximal Supersymmetry}, \hepth{0201081}.

\bibitem{GSW87}
M.\ B.\ Green, J.\ H.\ Schwarz and E.\ Witten, {\it Superstring
Theory, Vols 1 \& 2}, Cambridge University Press, 1987.

\bibitem{Metsaev01}
R.\ R.\ Metsaev, 
{\it Type IIB Green-Schwarz Superstring in Plane Wave 
Ramond-Ramond Background}, \npb{625}{2002}{70}, \hepth{0112044}.

\bibitem{BMN02}
D.\ Berenstein, J.\ M.\ Maldacena and H.\ Nastase,
{\it Strings in Flat Space and pp-waves from ${\cal N}=4$ 
Super Yang-Mills}, \jhep{04}{2002}{013}, \hepth{0202021}.

\bibitem{MST00}
J.\ McGreevy, L.\ Susskind and N.\ Toumbas,
{\it Invasion of the Giant Gravitons from Anti de Sitter Space},
\jhep{06}{2000}{008}, \hepth{0003075}.

\bibitem{Witten98}
E.\ Witten, {\it Baryons and Branes in Anti de Sitter Space},
\jhep{07}{1998}{006}, \hepth{9805112}.

\bibitem{ST02}
K.\ Skenderis and M.\ Taylor,
{\it Branes in AdS and pp-wave Spacetimes}, \hepth{0204054}.

\bibitem{KR01}
A.\ Karch and L.\ Randall, 
{\it Open and Closed String Interpretation of SUSY CFT's on 
Branes with Boundaries}, \jhep{06}{2001}{063}, \hepth{0105132}.

\bibitem{BHLN02}
V.\ Balasubramanian, M.\ Huang, T.\ S.\ Levi and A.\ Naqvi,
{\it Open Strings from $\caln=4$ Super Yang-Mills},
\hepth{0204196}.

\bibitem{RY98}
S.\ Rey and J.\ Yee, 
{\it Macroscopic Strings as Heavy Quarks: Large-N Gauge Theory and
Anti de Sitter Supergravity}, 
\epjd{22}{2001}{379}, \hepth{9803001}.

\bibitem{Maldacena98}
J.\ M.\ Maldacena, 
{\it Wilson Loops in Large N Field Theories},
\prl{80}{1998}{4859}, \hepth{9803002}.

\bibitem{Imamura98b}
Y.\ Imamura, 
{\it Supersymmetries and BPS Configurations on Anti-de Sitter Space}, 
\npb{537}{1999}{184}, \hepth{9807179}.

\bibitem{BISY98}
A.\ Brandhuber, N.\ Itzhaki, J.\ Sonnenschein and S.\ Yankielowicz,
{\it Baryons from Supergravity}, 
\jhep{07}{1998}{020}, \hepth{9806158}.

\bibitem{Imamura98a}
Y.\ Imamura, 
{\it Baryon Mass and Phase Transitions in Large N Gauge Theory},
\ptp{100}{1998}{1263}, \hepth{9806162}.

\bibitem{CGS98}
C.\ G.\ Callan, A.\ G\"uijosa and K.\ G.\ Savvidy,
{\it Baryons and String Creation from the Fivebrane Worldvolume
Action}, \npb{547}{1999}{127}, \hepth{9810092}.

\bibitem{CGMV99}
B.\ Craps, J.\ Gomis, D.\ Mateos and A.\ Van Proeyen,
{\it BPS Solutions of a D5-brane Worldvolume in a D3-brane 
Background from Superalgebras}, 
\jhep{04}{1999}{004}, \hepth{9901060}.

\bibitem{GRST99}
J.\ Gomis, A.\ Ramallo, J.\ Sim\'on and P.\ K.\ Townsend,
{\it Supersymmetric Baryonic Branes},
\jhep{11}{1999}{019}, \hepth{9907022}.

\bibitem{CM97}
C.\ G.\ Callan and J.\ M.\ Maldacena,
{\it Brane Dynamics From the Born-Infeld Action},
\npb{513}{1998}{198}, \hepth{9708147}.

\bibitem{Gibbons97}
G.\ W.\ Gibbons,
{\it Born-Infeld Particles and Dirichlet p-branes}, 
\npb{514}{1998}{603}, \hepth{9709027}.

\bibitem{Penrose76}
R.\ Penrose, {\it Any Spacetime Has a Plane-wave as a Limit},
in `Differential Geometry and Relativity', p. 271, 
Reidel, Dordrecht, 1976.

\bibitem{Guven00}
R.\ G\"uven, {\it Plane Wave Limits and T-Duality},
\plb{482}{2000}{255}, \hepth{0005061}.

\bibitem{BFP02}
M.\ Blau, J.\ Figueroa-O'Farrill and G.\ Papadopoulos,
{\it Penrose Limits, Supergravity and Brane Dynamics}, \hepth{0202111}.

\bibitem{LU97}
U.\ Lindstr\"om and R.\ von Unge, 
{\it A Picture of D-branes at Strong Coupling},
\plb{403}{1997}{233}, \hepth{9704051}.

\bibitem{BT98}
E.\ Bergshoeff and P.\ K.\ Townsend,
{\it Super D-branes Revisited}, \npb{531}{1998}{226}, \hepth{9804011}.

\bibitem{BKOP97}
E.\ Bergshoeff, R.\ Kallosh, T.\ Ort\'\i n and G.\ Papadopoulos,
{\it Kappa-symmetry, supersymmetry and intersecting branes},
\npb{502}{1997}{149}, \hepth{9705040}.

\bibitem{IKM02}
N.\ Itzhaki, I.\ R.\ Klebanov and S.\ Mukhi,
{\it PP Wave Limit and Enhanced Supersymmetry in Gauge Theories},
\jhep{03}{2002}{048}, \hepth{0202153}.

\bibitem{GO02}
J.\ Gomis and H.\ Ooguri, {\it Penrose Limit of N=1 Gauge Theories},
\hepth{0202157}.

\bibitem{PS02}
L.\ A.\ Pando Zayas and J.\ Sonnenschein,
{\it On Penrose Limits and Gauge Theories}, \hepth{0202186}.

\bibitem{GGT97}
J.\ P.\ Gauntlett, J.\ Gomis and  P.\ K.\ Townsend,
{\it BPS Bounds for Worldvolume Branes}, 
\jhep{01}{1998}{003}, \hepth{9711205}.

\bibitem{MNT02}
D.\ Mateos, S.\ Ng and P.\ K.\ Townsend,
{\it Supercurves}, \hepth{0204062}, to be published in {\it Phys.\
Lett.\ {\bf B}}.
 
\bibitem{MNT01}
D.\ Mateos, S.\ Ng and P.\ K.\ Townsend,
{\it Tachyons, Supertubes and Brane/Anti-Brane Systems}, 
\jhep{03}{2002}{016}, \hepth{0112054}.

\bibitem{MT01}
D.\ Mateos and P.\ K.\ Townsend, {\it Supertubes},
\prl{87}{2001}{011602}, \hepth{0103030}.

\bibitem{Seki02}
S.\ Seki, {\it D5-brane in Anti-de Sitter Space and Penrose Limit}, \hepth{0205266}.

\end{thebibliography}
\end{document}